\documentclass[aps,prx,groupedaddress,superscriptaddress,reprint,showpacs,pdftex,longbibliography]{revtex4-1}
\usepackage{graphicx}
\usepackage{tabularx}
\usepackage{multirow}
\usepackage{longtable}
\usepackage{dcolumn}
\usepackage{bm}
\usepackage{amsmath}
\usepackage{amssymb}
\usepackage{longtable}
\usepackage{txfonts}
\usepackage{url}
\usepackage[colorlinks=true, linkcolor=blue, anchorcolor=blue, citecolor=blue, urlcolor=magenta]{hyperref}
\usepackage[usenames,dvipsnames]{color}
\definecolor{Gray}{gray}{0.0}
\definecolor{lightGray}{gray}{0.35}

\begin{document}
\title{
(La,Th)H$_{10}$: the potential high-$T\rm_{c}$ superconductors stabilized thermodynamically below 200 GPa
}
\author{Peng Song}
\affiliation{School of Information Science, JAIST, Asahidai 1-1, Nomi, Ishikawa 923-1292, Japan}
\author{Zhufeng Hou}
\affiliation{State Key Laboratory of Structural Chemistry, Fujian Institute of Research on the Structure of Matter, Chinese Academy of Sciences, Fuzhou 350002, China}
\author{Kenta Hongo}
\affiliation{Research Center for Advanced Computing Infrastructure, JAIST, Asahidai 1-1, Nomi, Ishikawa 923-1292, Japan}
\author{Ryo Maezono}
\affiliation{School of Information Science, JAIST, Asahidai 1-1, Nomi, Ishikawa 923-1292, Japan}

\vspace{10mm}

\date{\today}
\begin{abstract}
The recent high-pressure experimental discovery of superconductivity in (La,Y)H$_{10}$, (La,Ce)H$_{9}$, (La,Ce)H$_{10}$, (Y,Ce)H$_{9}$, and (La,Nd)H$_{10}$ shows that the ternary rare-earth clathrate hydride can be promising candidate for high-temperature superconductor.
In this work, we theoretically demonstrate that the combination of actinide-metal thorium (Th) and rare-earth-metal lanthanum (La) with hydrogen can also form some ternary hydrides with cage-like structures to be stable at 200 GPa.
Using the evolutionary algorithms combined with the first-principles calculations, we have predicted the pressure-dependent ternary phase diagram of La$_{x}$Th$_{y}$H$_{z}$, 
particularly including the case of (La$_{1-x}$Th$_{x}$)H$_{n}$ [or designated as (La,Th)H$_{n}$ for simplicity].
Our calculations show that the hydrogen-rich phases such as (La,Th)H$_{9}$ (only including $P\bar{6}m2$-LaThH$_{18}$) and (La,Th)H$_{10}$ (including $I4/mmm$-La$_{3}$ThH$_{40}$, $R\bar{3}m$-LaThH$_{20}$, and $I4/mmm$-LaTh$_{3}$H$_{40}$) with H$_{29}$ and H$_{32}$ cages can be thermodynamically stable below 200 GPa.
However, the phase decomposition can happen to only (La,Th)H$_{9}$ when the pressure is above 150 GPa. More importantly, the electron-phonon coupling (EPC) calculations show that the (La,Th)H$_{10}$ series could the potential superconductors, of which $I4/mmm$-La$_{3}$ThH$_{40}$ at 200 GPa exhibits the large EPC constant $\lambda$ = 2.46 with a highest transition temperature ($T_\mathrm{c}$) of 210 K.
Since there are few previous studies on ternary hydrides composed of actinide metals, the present work would greatly stimulate the further discovery of this type of ternary hydrides and provide useful guidance for the high-pressure experimental studies on them.
\end{abstract}
\maketitle

\section{Introduction}
\label{sec.intro}
The current progress of the theoretical studies on high-pressure superconducting hydrides has reached an unprecedented stage,
with the resolution of almost all binary hydrides and some ternary and multiple hydrides.~\cite{2020FLO,2018LEV,2021SHI,2022LIL}
The superconducting transition temperatures ($T_\mathrm{c}$) of the H-cage-containing hydrides such as MgH$_{6}$, YH$_{6}$, YH$_{9}$, YH$_{10}$, LaH$_{10}$, Li$_{2}$MgH$_{16}$,
and LaYH$_{20}$ have been predicted to be close to or even greater than the room temperature.~\cite{2015FEN,2017PEN,2017LIU,2019SUN,2021SEM}
The near-room-temperature superconductivity in lanthanum hydride and lanthanum-yttrium ternary hydrides has been demonstrated in high-pressure experiments with the XRD diffraction
measurements to confirm their crystal structures.~\cite{2019SOM,2019DRO,2021SEM}

\vspace{2mm}
Most of the high-temperature superconducting hydrides synthesized currently in experiment are crystalized
in the hydrogen clathrate structure composed of alkaline-earth and rare-earth metals.~\cite{2021SEM,2019SOM,2019DRO,2019SAL,2020SEM,2021TRO,2021KON,2022MA,2022LI,2021SHA,2021CHE}
The presence of hydrogen in the form of cage structures, which could potentially reduce the product of pressure and volume ($PV$) for its contribution to enthalpy, is one of the fundamental factors to stabilize these clathrates at lower pressure. Thanks to the significant H-derived electron density of states at the Fermi level and the
robust electron-phonon coupling associated with the weak bonding of H atoms inside the cage, the cage
structures of metal hydrides have been predicted to exhibit potential high-temperature superconductivity.~\cite{2012WAN,2017PEN,2019SUN}
For the rare-earth clathrates that have been predicted by theoretical calculations to have superconductivity, not all of them have been verified by
high-pressure measurements. One of the plausible reasons may be ascribed to the significant magnetic characteristics.~\cite{2017PEN,2019PE,2021MA,2020ZHO,2020SEMb}
The noteworthy thing is that high-temperature superconductivity can be preserved in the sodalite-like structures, which could be formed by substituting carbon and/or nitrogen for hydrogen in clathrates, as reported in very recent theoretical calculations.~\cite{2022HAI,2022SAN,2022DFJ,2022ZLY,2022DSS}
Besides, a thermodynamically stable carbon-boron sp$^{3}$-bonded clathrate determined by particle-swarm structure prediction approach was successfully synthesized by Zhu \textit{et al.}~\cite{2020ZHU}
These studies illustrate that the high-temperature superconductivity in metal hydrides exhibits a strong correlation with the clathrate structure.

\vspace{2mm}
Recently, the superconductivity of ternary clathrate hydride has also attracted a lot of attention, and most of them cover the hydrides of alkaline-earth metals and rare-earth metals.~\cite{2021SEM,2019LIA,2021SONa,2020SUK,2021SHIb,2022CHE,2022BI,2022SONb,2022WAN,2022SONc,2022SONd,2022SEM}
Although the theoretically predicted $T_\mathrm{c}$ of Li$_{2}$MgH$_{16}$ composed of alkali and alkaline-earth metals at 250 GPa is 473 K, its enthalpy of formation lies above the convex hull~\cite{2019SUN,2020WAN} and its synthesis in experiment is still a great challenge.
It is noteworthy to mention that partial replacement of La atoms by magnetic Nd atoms results in significant suppression of superconductivity in LaH$_{10}$.~\cite{2022SEM}
This provides a hint to design new superconducting materials. Among the ternary hydrides with cage structures, we note that some of their specific compositions such as (A,B)H$_{6}$, (A,B)H$_{9}$, and (A,B)H$_{10}$ can be obtained by mixing their parent compounds (AH$_{x}$ and BH$_{x}$, $x$ = 6, 9, and 10) with the same cage structures.~\cite{2019LIA,2022SONb,2021SEM,2022CHE,2022BI}
The ground state electron configurations of La and Th are [Xe]5d$^{1}$6s$^{2}$ and [Rn]5d$^{2}$7s$^{2}$. Their  electronegativity and atomic radius are quite close to each other. More interestingly, both of them can form the stable binary hydride $Fm\bar{3}m$-AH$_{10}$ at pressure lower than 200 GPa,~\cite{2017LIU,2018KVA}
which have also been confirmed in experiment.~\cite{2019SOM,2019DRO,2020SEM} It is natural for us to consider whether (La,Th)H$_{10}$ could be stable or not. If so, the chemical synthesis of (La,Th)H$_{10}$  would be straightforward and also similar to that of (La,Y)H$_{10}$~\cite{2021SEM}.
Therefore, we concentrate on the ternary compounds of La-Th-H and their structural stability as well as superconductivity.

\vspace{2mm}
In this work, we have systematically investigated the stability and superconductivity of La$_{x}$Th$_{y}$H$_{z}$ under high pressure using the first-principles calculations combined with the evolutionary algorithms for structure search.
The hydrogen-rich phases such as $I4/mmm$-La$_{3}$ThH$_{40}$, $R\bar{3}m$-LaThH$_{20}$, and $I4/mmm$-LaTh$_{3}$H$_{40}$ are found to be stabilized below 200 GPa and all of them exhibit potential high-temperature superconductivity. In particular, $I4/mmm$-La$_{3}$ThH$_{40}$ at 200 GPa is predicted to have a highest $T_\mathrm{c}$ of 210 K.
\section{Method}
The structure search for La$_{x}$Th$_{y}$H$_{z}$ was performed using the evolutionary variable-composition simulation as implemented in the USPEX (Universal structure predictor: evolutionary Xtallography) software~\cite{2006GLA,2013LYA}. The crystal structures of La$_{x}$Th$_{y}$H$_{z}$ were considered with a maximum number of 24 atoms per cell at a series of fixed pressure including 5, 50, 100, and 200 GPa.
200 structures in the first generation were created randomly, while 100 structures in the subsequent every generation were obtained by four different manipulations, namely, 40\% by hereditary, 40\% by random creation, 10\% by mutation, and 10\% by soft mutation. A total of 100 generations were iterated for each considered composition of La$_{x}$Th$_{y}$H$_{z}$.
Since the high-temperature superconductivity is usually observed in hydrogen-rich materials, herein we omitted the composition of $(x + y)/z < 1$ by the seed technique~\cite{2013LYA} in USPEX.
To further check whether the stable phase may have a maximum number of more than 24 atoms per cell or not, we carried out a specific structure search for the (LaH$_{10}$)$_{x^{\prime}}$(ThH$_{10}$)$_{y^{\prime}}$ system under the pressure of 100 and 200 GPa.
Each structure underwent a four-round optimization by the first-principles calculations, which were carried out using the VASP (Vienna \textit{ab initio} software package)~\cite{1993KRE,1994KRE,1996KRE,1996KREb} code. A force convergence criterion of 0.02 eV/\AA~was specified in the last round of optimization. The interaction between ions and electrons was described by the projector augmented waves (PAW) method~\cite{1994BL,1999KRE} and the recommended PAW potentials for La, Th, and H were employed. The cutoff energy of 600 eV was used for the plane-wave basis-set. The Perdew-Burke-Ernzerhof (PBE) exchange-correlation functional~\cite{1996PER} was employed.
The \textit{k}-points in the first Brillouin zone (BZ) were sampled with a setup of 0.1 \AA$^{-1}$ for the smallest allowed spacing between \textit{k}-points.

\vspace{2mm}
Based on the enthalpies of La$_{x}$Th$_{y}$H$_{z}$ obtained by the VASP calculations, the convex hull construction for the La$_{x}$Th$_{y}$H$_{z}$ phase diagram was determined using the Pymatgen~\cite{2013ONG} tool.
The enthalpy of a phase above the convex hull indicates that such a phase is unstable and thus it would be decomposed into other phases.

\vspace{2mm}
For the stable phases of La$_{x}$Th$_{y}$H$_{z}$, their phonon dispersions and superconductivity were calculated within the density functional perturbation theory (DFPT)~\cite{2001BAR} as implemented in Quantum ESPRESSO (QE)~\cite{2009GIA} code. The plane wave expansion of the electron eigenstates had a cutoff of 113 Ry. In particular, the PAW potenitals of La and Th were generated according to the valence configurations of 4f$^{0}$5d$^{1}$6s$^{2}$6p$^{0}$ and 6d$^{1}$7s$^{2}$7p$^{0}$5f$^{1}$, respectively. The PBE functional~\cite{1996PER} was used too.
The superconducting transition temperature $T_\mathrm{c}$ is evaluated by the Allen-Dynes-modified McMillan formula~\cite{1975ALL}:
\begin{equation}
T_\mathrm{c} = \frac{\omega_{\rm{log}} f_{1} f_{2}}{1.2} \rm{exp}\left(\frac{-1.04 (1 + \lambda)}{\lambda (1 - 0.62\mu^{\ast}) - \mu^{\ast}}\right),
\end{equation}
with
\begin{eqnarray}
\nonumber 
  f_{1}f_{2} &=& \sqrt[3]{1 + \left[\frac{\lambda}{2.46 (1 + 3.8 \mu^{\ast}))}\right]^{\frac{3}{2}}} \\
    & & \times \left[1 - \frac{\lambda^{2}(1 - \omega_{2}/\omega_{\rm{log}})}{\lambda^{2} + 3.312(1 + 6.3\mu^{\ast})^{2}}\right],
\end{eqnarray}
where $\mu^{\ast}$ is the Coulomb pseudopotential parameter. The widely accepted value of 0.1 for $\mu^{\ast}$ is used herein. The electron-phonon coupling constant $\lambda$, logarithmic average phonon frequency $\omega_{\rm{log}}$, and mean square frequency $\omega_2$ are defined as below
\begin{equation}
\label{eq:epc}
\lambda = 2 \int{\frac{\alpha^{2}F(\omega)}{\omega}}d{\omega},
\end{equation}
\begin{equation}
\label{eq:wlog}
\omega_{\rm{log}} = {\rm{exp}}\left[ \frac{2}{\lambda}\int{\frac{d\omega}{\omega}\alpha^{2}F(\omega){\rm{log}}\omega}      \right ],
\end{equation}
and
\begin{equation}
\label{eq:omega2}
\omega_{2}  = \sqrt{\frac{1}{\lambda} \int{\left[\frac{2\alpha^{2}F(\omega)}{\omega} \right] \omega^{2}d\omega }},
\end{equation}
respectively. In the case of $f_{1}f_{2}$ = 1, the Allen-Dynes-modified McMillan formula is restored to the original McMillan formula~\cite{1968MCM}.

\section{Results}
\label{sec.results}
\subsection{Thermodynamic Stability and crystal structure of La-Th-H system}
\begin{figure*}[htbp]
  \begin{center}
    \includegraphics[width=\linewidth]{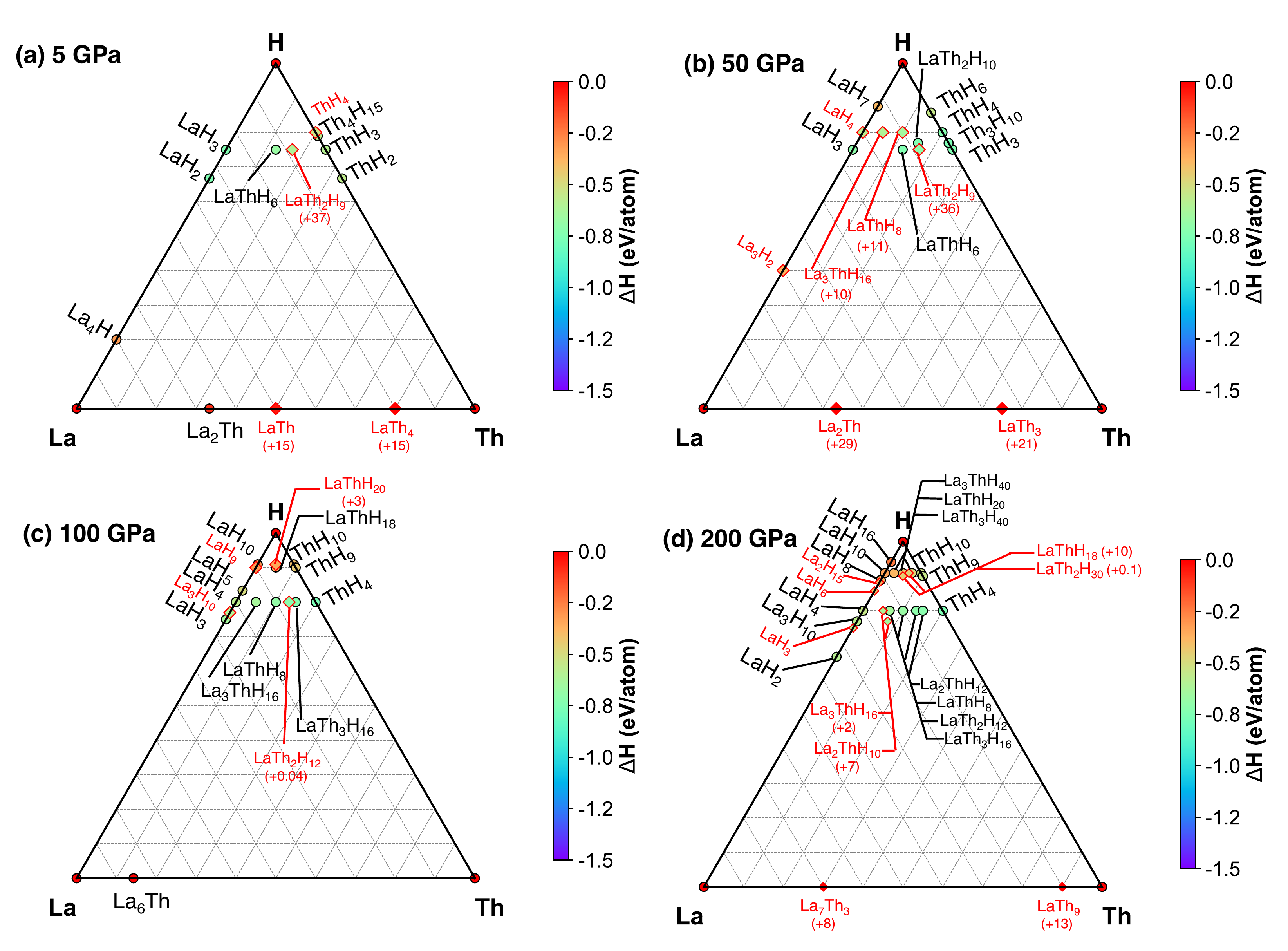}
    \caption{Ternary convex hull of La$_{x}$Th$_{y}$H$_{z}$ at the pressure of 5, 50, 100, 200 GPa. Thermodynamically stable and metastable phases are indicated by the circle and red-edge square symbols, respectively. For the metastable ternary phases, their specific values (in unit of meV/atom) of the heat of formation above the hull are also remarked in parenthesis.}
    \label{fig.convex_hull}
  \end{center}
\end{figure*}

The convex hulls for the stable and metalstable phases of La$_{x}$Th$_{y}$H$_{z}$ at the pressure of 5, 50, 100, and 200 GPa are presented in Fig.~\ref{fig.convex_hull}. The complete information about the enthalpy of formation and structure properties of these phases is listed in Table S1-S5 in the Supporting Information (SI). There is no stable ternary phase in the phase diagram of La$_{x}$Th$_{y}$H$_{z}$ at ambient pressure, which is easily accessible through the Materials Project (MP) database~\cite{2013JAI,2008ONG}.
The binary hydrides of both La and Th at ambient pressure have been extensively studied in experiment~\cite{1989BOR} and theoretical calculations~\cite{1997WAN,2007SHE}. It is worth noting that Th$_{4}$H$_{15}$ has been synthesized and also found to be a superconductor at low pressure (5 GPa).~\cite{2021WANb}
Our calculations show that at 5 GPa there is a stable ternary phase, i.e., $Pm\bar{3}$-LaThH$_{6}$.
For the Th-H binary phase, our calculations predict a new stable phase, i.e., $R\bar{3}c$-ThH$_{3}$, with a lower energy than $\frac{3}{7}$ThH$_{2}$ + $\frac{1}{7}$Th$_{4}$H$_{15}$, which has not been found in previous study~\cite{2018KVA}.
At 50 GPa, our predicted binary phases are in overall agreement with those reported in previous studies~\cite{2018KVA,2020KRU,2017PEN}. For the ternary case, the low-symmetry $Cm$-LaTh$_{2}$H$_{10}$ is found to be also stabilized at 50 GPa beside $Pm\bar{3}$-LaThH$_{6}$. At 100 GPa, $Pm\bar{3}$-LaThH$_{6}$ becomes unstable, while more new stable phases appear along (LaH$_{4}$)$_{x^{\prime}}$(ThH$_{4}$)$_{y^{\prime}}$ line.
We should point out that the $Fmmm$-LaTh$_{2}$H$_{12}$ phase is just located at 0.04 meV/atom above the convex hull, which could also be stabilized owing to such a tiny energy.
Although LaH$_{9}$ is a metastable phase at 100 GPa, $P\bar{6}m2$-LaThH$_{18}$ is a stable ternary hydrogen-rich phase emerging along (LaH$_{9}$)$_{x^{\prime}}$(ThH$_{9}$)$_{y^{\prime}}$.
When the pressure is further increased to 200 GPa, most of the stable phases with compositions along (LaH$_{4}$)$_{x^{\prime}}$(ThH$_{4}$)$_{y^{\prime}}$ at 100 GPa still remain on the convex hull except $Pmmm$-La$_{3}$ThH$_{16}$. Phase decomposition happens to $P\bar{6}m2$-LaThH$_{18}$ at 200 GPa. 
Supplementary structure search at 200 GPa was carried out specifically for the compositions along (LaH$_{10}$)$_{x^{\prime}}$(ThH$_{10}$)$_{y^{\prime}}$ line. In this way, several novel hydrogen-rich phases such as $I4/mmm$-La$_{3}$ThH$_{40}$, $R\bar{3}m$-LaThH$_{20}$, and $I4/mmm$-LaTh$_{3}$H$_{40}$ were found to be thermodynamically stable.
In addition, $Immm$-LaTh$_{2}$H$_{30}$ is located at 0.2 meV/atom above the convex hull and thus it is also extremely close to stable phase.

\begin{figure*}[htbp]
  \begin{center}
    \includegraphics[width=\linewidth]{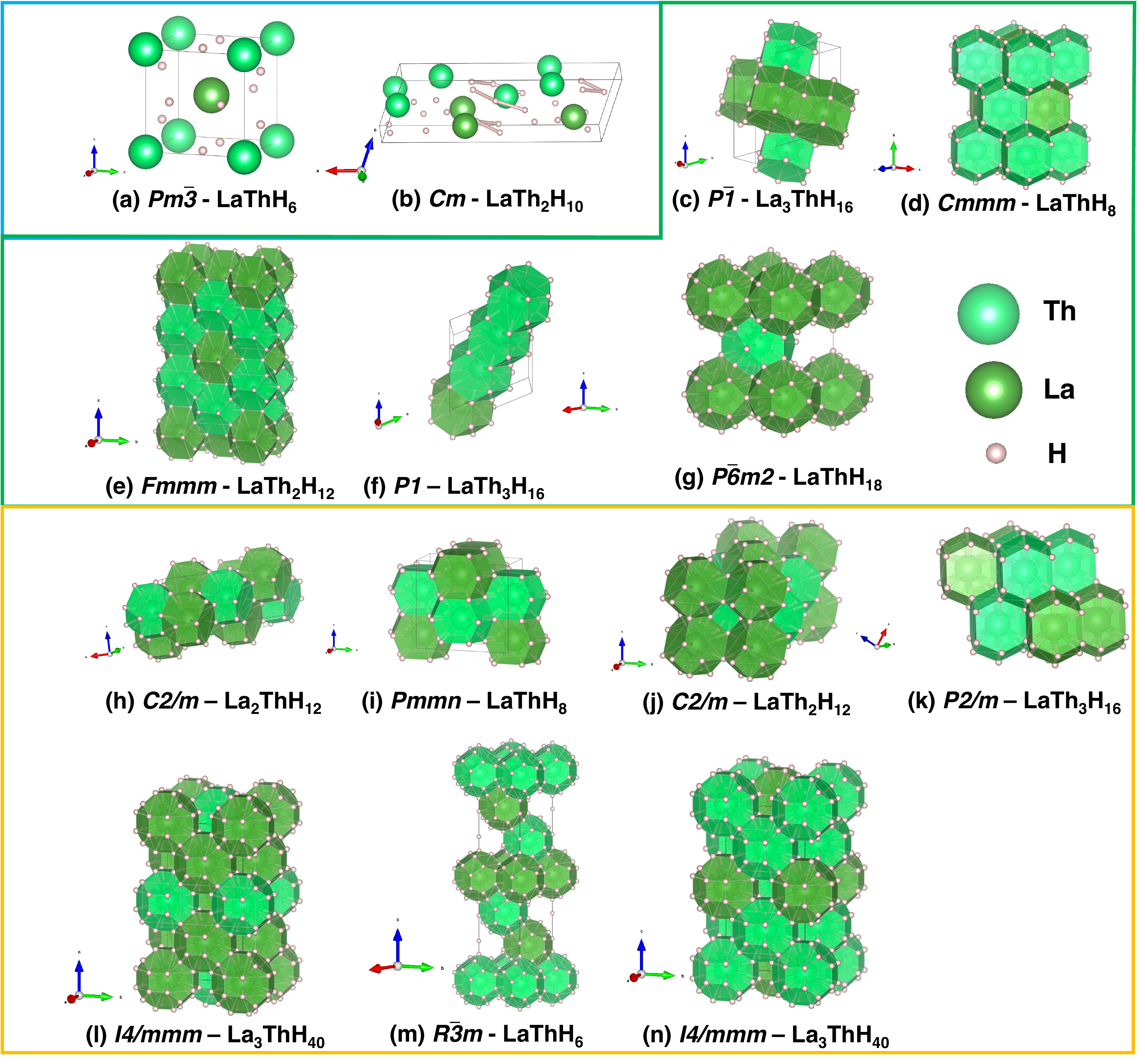}
    \caption{Crystal structures of La$_{x}$Th$_{y}$H$_{z}$. The crystal structures in (a)-(b) are stable in the pressure range over 5 to 50 GPa, (c)-(g) are stable at 100 GPa, and (h)-(o) are stable at 200 GPa. (La,Th)H$_{4}$ [(c)-(f),(h)-(k)], (La,Th)H$_{9}$ [(g)], and (La,Th)H$_{10}$ [(l)-(n)] are clathrate structures composed of H$_{18}$, H$_{29}$, and H$_{32}$ cages, respectively, which are represented by the polyhedrons.}
    \label{fig.structure}
  \end{center}
\end{figure*}

\vspace{2mm}
The crystal structures for the predicted stable phases of La$_{x}$Th$_{y}$H$_{z}$ are shown in Fig.~\ref{fig.structure}.
Except for $Pm\bar{3}$-LaThH$_{6}$ and $Cm$-LaTh$_{2}$H$_{10}$, which are stable below 50 GPa, the rest of the predicted stable phases of (La,Th)H$_{4}$, (La,Th)$_{9}$, and (La,Th)H$_{10}$ take the clathrate structures composed of H$_{18}$, H$_{29}$, and H$_{32}$ cages, respectively. The H$_{18}$, H$_{29}$, and H$_{32}$ cages consist of a dodecahedron (8 eight quadrangle and 4 hexagonal faces), a dodecahedron (12 pentagonal and 6 quadrangle faces), and an octahedron (12 hexagonal and 6 quadrangle faces), respectively.
The similar clathrate structures have been extensively studied in hydrogen storage materials~\cite{2007STR}. The binary hydrides with clathrate structures have been predicted to possess excellent superconducting properties.~\cite{2017PEN,2018KVA}
Although the (La,Th)H$_{4}$ with several different compositions (i.e., the La/Th ratios) would undergo phase transition from 100 GPa to 200 GPa, the same H$_{18}$ cage structure is kept during the pressure-induced phase transition. This is because these cage structures are more competitively stable under high pressure because of the relatively reduced contribution of $PV$ term to enthalpy.~\cite{2017PEN}

\vspace{2mm}
From ambient pressure to high pressure, the thermodynamically stable phases of La$_{x}$Th$_{y}$H$_{z}$ are gathered mainly on four lines, i.e., (La,Th)H$_{3}$, (La,Th)H$_{4}$, (La,Th)H$_{9}$, and (La,Th)H$_{10}$), as seen from Fig.~\ref{fig.structure}.
The shortest decomposition path of LaThH$_{6}$ and LaThH$_{18}$ was determined using the Pymatgen~\cite{2013ONG} tool.
The decomposition reaction of LaThH$_{6}$ at 100 GPa is LaThH$_{6}$ $\rightarrow$ $\frac{1}{3}$La + ThH$_{4}$ + $\frac{2}{3}$LaH$_{3}$, and hence the applying more high pressure to LaThH$_{6}$ would not obtain a new ternary phase.
Similarly, the decomposition reaction of LaThH$_{18}$ at 200 GPa is LaThH$_{18}$ $\rightarrow$ $\frac{1}{6}$LaTh$_{3}$H$_{40}$ + $\frac{1}{2}$LaThH$_{20}$ + $\frac{1}{3}$LaH$_{4}$.
From the two main contributions (i.e., internal energy and $PV$) to the pressure-dependent relative enthalpy of LaThH$_{18}$ given with respect to the decomposed phases, as shown in Fig.~S1 in the SI, it is found that the contribution from $PV$ plays a more significant role in the pressure-dependent relative enthalpy, while the one from internal energy almost remains unchanged. 

\vspace{2mm}
For the newly predicted ternary phases in the pressure range of their thermodynamical stabilization, we have checked their electronic properties including the electronic band structure and partial density of states (PDOS). The results are presented in Fig.~S2 in the SI. All of these studied La$_{x}$Th$_{y}$H$_{z}$ phases exhibit metallic behavior.
The electronic states at the Fermi level ($E_\mathrm{F}$) of these phases are strongly dependent on the hydrogen content.
The contribution of H atoms to the electronic states at $E_\mathrm{F}$ is nearly zero in LaThH$_{6}$ and LaTh$_{2}$H$_{10}$, and it is also negligible in the (La,Th)H$_{4}$ system, suggesting that the electronic structures of these phases may be manifested mainly through La-Th. A previous experiment study showed that the observed $T_\mathrm{c}$ of the La-Th alloy (i.e., $T_\mathrm{c}$ below 6 K) decreased with increasing Th content.~\cite{1973SAT}
This implies that high-temperature superconductivity may not be achieved in the aforementioned phases with lower H content.
It is worth noting that the hydrogen fraction of the total DOS at $E_\mathrm{F}$ (denoted as H$_\mathrm{DOS}$) is high in (La,Th)H$_{9}$ and (La,Th)H$_{10}$.
According to Belli \textit{et al.}~\cite{2021BEL}, superconducting $T_\mathrm{c}$ in hydrides maintains a strong positive correlation with both H$_\mathrm{DOS}$ and electron localization function (ELF) value.
According to this argument, (La,Th)H$_{9}$ and (La,Th)H$_{10}$ might be potential high-temperature superconductors and more detailed results are given in next subsection.

\subsection{Phonon dispersion and superconductivity of newly predicted La$_{x}$Th$_{y}$H$_{z}$ phase under the harmonic approximation}
We further examine the lattice-dynamic stability and superconductivity of these thermodynamically stable phases in the studied pressure range (5-200 GPa) by performing the phonon and electron-phonon coupling (EPC) calculations under the harmonic approximation.
It is found that $Fmmm$-LaTh$_{2}$H$_{12}$, $P\bar{6}m2$-LaThH$_{18}$, $Pmmn$-LaThH$_{8}$, and $P2/m$-LaTh$_{3}$H$_{16,}$ exhibit phonon with imaginary frequencies, indicting that they are dynamically unstable, while the remaining thermodynamically stable phases do not.
For the phases that meet both thermodynamic stability and lattice-dynamic stability, we further calculated their Eliashberg phonon spectral function $\alpha^{2}F$.
By integrating $\alpha^{2}F$, we obtained the EPC constant and then evaluated the $T_\mathrm{c}$ values. The main results are summarized in Table~\ref{table.Pdep}.
For the phase with lower H content, the (La,Th)H$_{3}$ and (La,Th)H$_{4}$ systems are unable to attain high temperature superconductivity due to their low EPC constants.
Herein we take $Pm\bar{3}$-LaThH$_{6}$ as an example. The EPC constant of $Pm\bar{3}$-LaThH$_{6}$ is 0.23 and its contribution from the vibration modes of H atoms is just 0.09 (i.e., accounting for 39\%).
The low-symmetry phase $P1$-La$_{3}$ThH$_{16}$ with the highest $T\rm_{c}$ among the (La,Th)H$_{3}$ and (La,Th)H$_{4}$ systems has a significant increase in H$_\mathrm{DOS}$ and its EPC constant ($\lambda$ = 0.68), of which the contribution from the vibrational modes of H atoms (i.e., accounting for 70\%).
For the phases with much higher H content, the contribution of H atoms to the electronic density of states at $E_\mathrm{F}$ is dominant, so we next focus on the superconductivity of three hydrogen-rich phases, namely, $I4/mmm$-La$_{3}$ThH$_{40}$, $R\bar{3}m$-LaThH$_{20}$, and $I4/mmm$-LaTh$_{3}$H$_{40}$.

\begin{figure*}[htbp]
  \begin{center}
    \includegraphics[width=\linewidth]{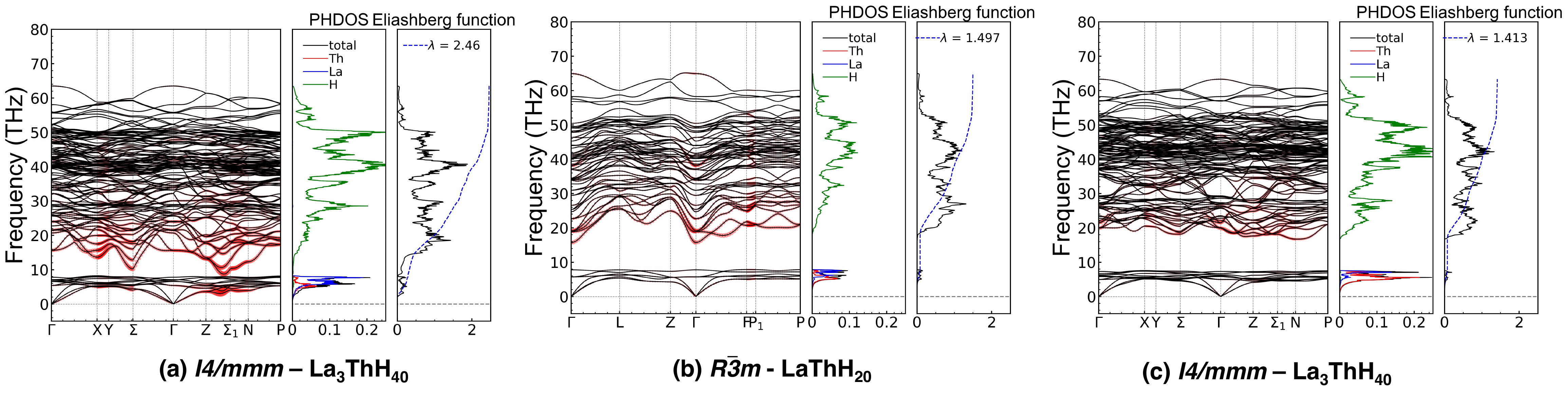}
    \caption{Phonon dispersions with the mode-resolved electron-phonon coupling (EPC) constant $\lambda_{q\nu}$ (indicated by the size of red circle), atom-projected and total phonon density of states (PHDOS), and Eliashberg spectral function of (a) $I4/mmm$-La$_{3}$ThH$_{40}$, (b) $R\bar{3}m$-LaThH$_{20}$, and (c) $I4/mmm$-LaTh$_{3}$H$_{40}$ at 200 GPa.}
    \label{fig.phonon}
  \end{center}
\end{figure*}

\vspace{2mm}
Figure~\ref{fig.phonon} shows the phonon dispersion with a mode-revolved EPC constant $\lambda_{q\nu}$, phonon density of states (PHDOS), and electron-phonon Eliashberg spectral function $\alpha^{2}F(\omega)$ for $I4/mmm$-La$_{3}$ThH$_{40}$, $R\bar{3}m$-LaThH$_{20}$, and $I4/mmm$-LaTh$_{3}$H$_{40}$.
The EPC constants of these three phases are 2.46, 1.50, and 1.41, respectively, which follow the similar descending order for the density of states at the Fermi level ($N_{E_\mathrm{F}}$).
The vibrational modes of phonons of these three phases can be clearly grouped into two regions according to the frequencies. The first region with phonon frequencies of 0-10 THz (denoted as region I) is dominated by the vibration modes of La and Th atoms.
For the contribution of phonon in this region to the EPC constants, it is about 0.07 in both LaThH$_{20}$ and LaTh$_{3}$H$_{40}$, while it is high up to 0.33 in  La$_{3}$ThH$_{40}$.
As seen from the mode-revolved EPC $\lambda_{q\nu}$, La$_{3}$ThH$_{40}$ shows a significant enhancement in EPC along the path of Z-$\rm{\Sigma_{1}}$-N in its first BZ. 
The second region with phonon frequencies of $\ge$ 10 Thz (denoted as region II) arise from the vibrational modes of H atoms. The contributions of this region in the EPC constant are 2.12 (86.5\%), 1.41 (94.8\%), and 1.34 (94.6\%), respectively.
The phonon vibrational modes of La$_{3}$ThH$_{40}$ exhibit significant softening in the region II especially at the $\Sigma$ point and in the path of Z-$\Sigma_{1}$-N.
It can be seen from $\lambda_{q\nu}$ that the softened phonon modes with frequencies of 10-20 Thz for La$_{3}$ThH$_{40}$ provides a large contribution (about 0.66) to the EPC constant.
Therefore, the difference in the EPC constants of these three hydrogen-rich phases is mainly caused by the appearance of the softened modes in La$_{3}$ThH$_{40}$.
Based on the Allen-Dynes formula, the highest $T_\mathrm{c}$ value is predicted to $I4/mmm$-La$_{3}$ThH$_{40}$ at 200 GPa, namely, $T_\mathrm{c}\simeq$ 210 K when the typical Coulomb pseudopotential parameter $\mu^{\ast}$ is taken as 0.1. The predicted $T_\mathrm{c}$ values of La$_{3}$ThH$_{40}$, LaThH$_{20}$, and LaTh$_{3}$Th$_{40}$ via the Allen-Dynes formula are slightly higher than the respective ones obtained by the McMillan formula.
For LaH$_{10}$ and ThH$_{10}$ as the parent compounds of these three superconducting ternary phases, their predicted $T_{c}$ values with the correction using a strong coupling factor $f_{1}f_{2}$ are much close to the experimentally measured ones.
$I4/mmm$-La$_{3}$ThH$_{40}$ possesses a higher EPC constant than its parent compounds, and thus its predicted $T_\mathrm{c}$ value with the same correction is significantly lifted.


\section{Conclusion}
\label{sec.conc}
In summary, we have employed the evolutionary algorithms and the first-principles calculations to explore the ternary phase diagram of the La-Th-H system in the pressure range over 5 to 200 GPa. It was found that the hydrogen-rich materials (La,Th)H$_{9}$ (only including LaThH$_{18}$) and (La,Th)H$_{10}$ (including La$_{3}$ThH$_{40}$, LaThH$_{20}$, and LaTh$_{3}$H$_{40}$) are thermodynamically stable below 200 GPa.
When the pressure is above 150 GPa, LaThH$_{18}$ would decompose into LaTh$_{3}$H$_{40}$, LaThH$_{20}$, and LaH$_{4}$.
La$_{3}$ThH$_{40}$ with the space group of $I4/mmm$ is predicted to have a largest EPC constant (i.e., $\lambda$ = 2.46) at 200 GPa among these stable phases of La-Th-H and thus to obtain a maximum superconducting transition temperature of 210 K.
The ternary clathrate hydrides of La-Ce-H, Y-Ce-H, and La-Y-H have already been synthesized in high-pressure experiments. La-Th-H exhibits the similar structural properties to the aforementioned three systems, suggesting that the La-Th-H system would be a promising ideal candidate to discover new high-temperature superconductor in high-pressure experiment.

\section*{Acknowledgments}
The computations in this work have been performed
using the facilities of
Research Center for Advanced Computing
Infrastructure (RCACI) at JAIST.
K.H. is grateful for financial support from
the HPCI System Research Project (Project ID: hp190169) and
MEXT-KAKENHI (JP16H06439, JP17K17762, JP19K05029, and JP19H05169).
R.M. is grateful for financial supports from
MEXT-KAKENHI (19H04692 and 16KK0097),
FLAGSHIP2020 (project nos. hp1
90169 and hp190167 at K-computer),
Toyota Motor Corporation, I-O DATA Foundation,
the Air Force Office of Scientific Research
(AFOSR-AOARD/FA2386-17-1-4049;FA2386-19-1-4015),
and JSPS Bilateral Joint Projects (with India DST).

\bibliographystyle{apsrev4-1}
\bibliography{references}


\clearpage
\newpage
\begin{table*}[htbp]
 \begin{center}
   \caption{Superconducting transition temperature ($T_\mathrm{c}$, in K) of La$_{x}$Th$_{y}$H$_{z}$ at different pressure ($P$, in GPa) evaluated according to the Allen-Dynes-modified McMillan formula (abbreviated to AD)~\cite{1975ALL} and the original McMillan formula (abbreviated to McM)~\cite{1968MCM}. The electron-phonon coupling constant ($\lambda$), logarithmic average phonon frequency ($\omega_\mathrm{log}$, in K), and density of states at Fermi level ($N_{E_\mathrm{F}}$, in states/spin/Ry/f.u.) are also listed.}
\label{table.Pdep}
\begin{tabular}{lccccccc}
\hline
\hline
Phase &Space group & $P$ & $\lambda$ & $\omega_{\rm log}$ & $N_{E_\mathrm{F}}$  &$T_\mathrm{c}$~(K) & $T_\mathrm{c}$ (K)  \\
 &   &~(GPa)           &   & ~(K)   & (states/spin/Ry/f.u. ) & (McM) & (AD)\\
 \hline
LaThH$_{6}$ & $Pm\bar{3}$  & 5        & 0.228    & 474.29& 11.87 &  $\ll$ 0.1 & $\ll$ 0.1 \\
 &   & 50        & 0.232    & 450.33& 9.17 &  $\ll$ 0.1 & $\ll$ 0.1 \\
  \hline
LaTh$_{2}$H$_{10}$ & $Cm$  & 50        & 0.223    & 504.28& 10.38 &  $\ll$ 0.1 & $\ll$ 0.1 \\
  \hline
La$_{3}$ThH$_{16}$ & $P\bar{1}$  & 100        & 0.680    & 1032.04& 14.30 &  33.4 & 34.7 \\
  \hline
La$_{2}$ThH$_{12}$ & $C2/m$  & 200        & 0.456    & 1282.81& 9.49 &  10.5 & 10.7 \\
  \hline
LaThH$_{8}$ & $Cmmm$  & 100        & 0.464    & 1118.00& 7.21 &  9.9 & 10.1 \\
  \hline
LaTh$_{2}$H$_{12}$ & $C2/m$  & 200        &0.316     & 1242.22 & 10.25 &  1.0 & 1.0 \\
  \hline
LaTh$_{3}$H$_{16}$ & $P1$  & 100        &0.40     &671.07  &  15.45 &2.9  &3.0  \\
  \hline
La$_{3}$ThH$_{40}$ & $I4/mmm$  & 200        &2.46     &998.62  &  20.55 &162.9   &210.0  \\
  \hline
LaThH$_{20}$ & $R\bar{3}m$  & 200        &1.50     &1414.83  & 8.92 &160.5   &179.2  \\
  \hline
LaTh$_{3}$H$_{40}$ & $I4/mmm$  & 200        &1.41     &1365.11  & 16.52 &147.4   & 163.8 \\

 \hline
 \hline
\end{tabular}
 \end{center}
\end{table*}
\clearpage
\newpage

\end{document}


\section{Supplemental Figure}
\label{supp_figure}
\begin{figure*}[htbp]
  \begin{center}
    \includegraphics[width=\linewidth]{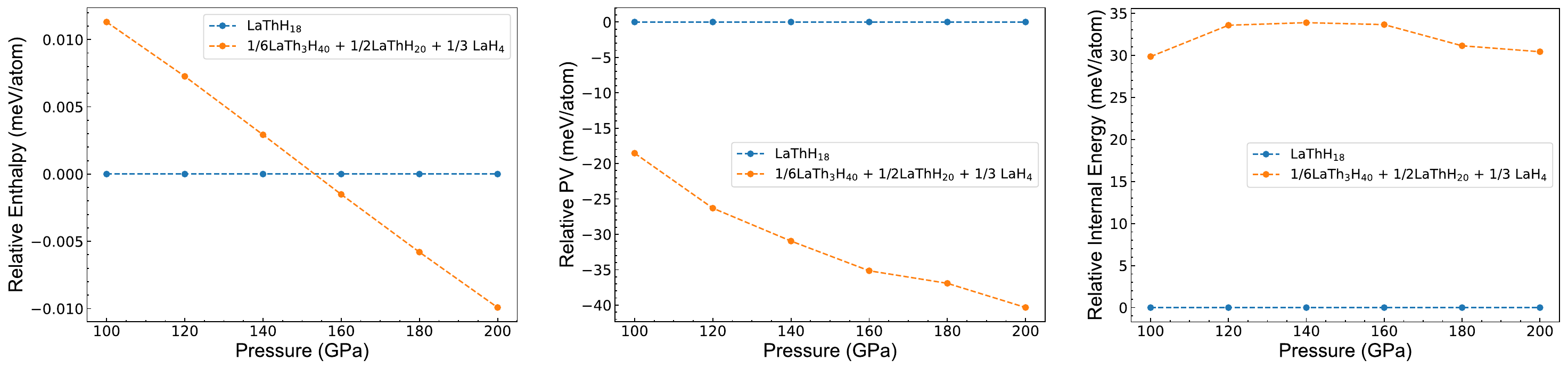}
    \caption{Pressure-dependent relative enthalpy of LaThH$_{18}$ with respect to its decomposed phases at 200 GPa.}
    \label{fig.compare}
  \end{center}
\end{figure*}
\begin{figure*}[htbp]
  \begin{center}
    \includegraphics[width=\linewidth]{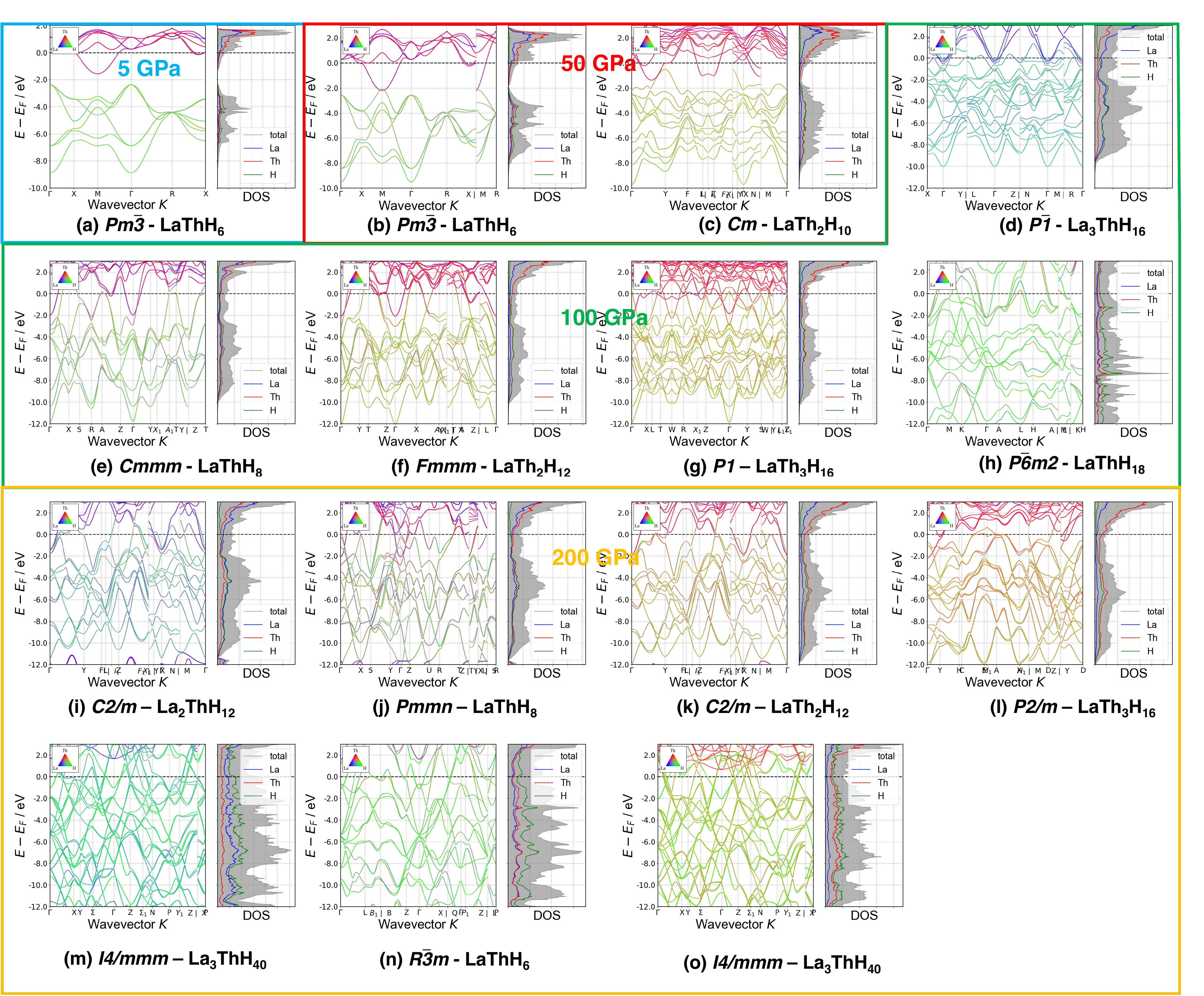}
    \caption{Electronic band structure and density of states (DOS) of newly predicted stable ternary phases of La$_{x}$Th$_{y}$H$_{z}$.}
    \label{fig.band_structure}
  \end{center}
\end{figure*}
\begin{figure*}[htbp]
  \begin{center}
    \includegraphics[width=\linewidth]{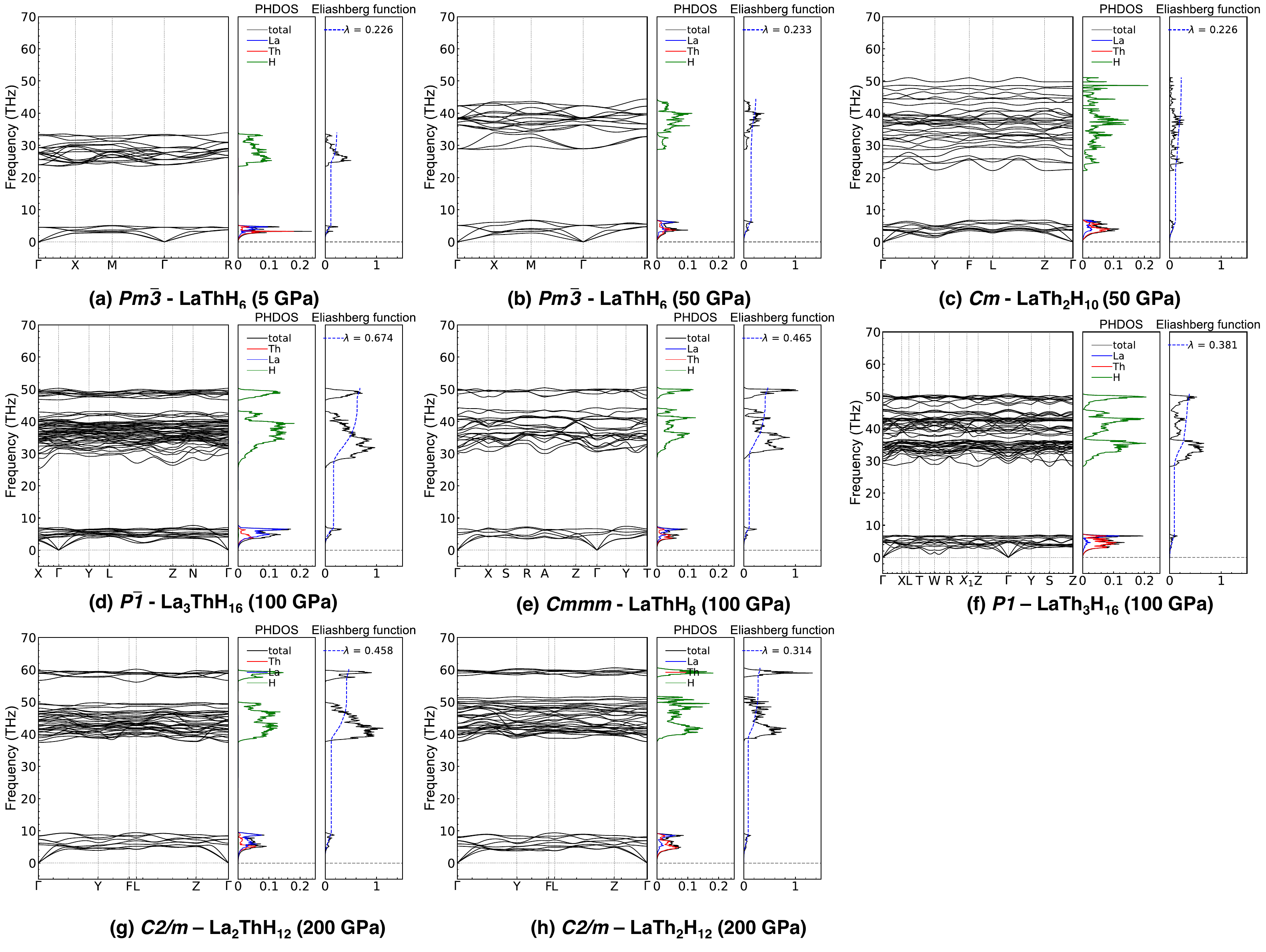}
    \caption{Phonon dispersion, phonon density of states(PHDOS), and Eliashberg spectral function of newly predicted stable ternary phases of La$_{x}$Th$_{y}$H$_{z}$ without imaginary modes.  }
    \label{fig.band_structure}
  \end{center}
\end{figure*}
\begin{figure*}[htbp]
  \begin{center}
    \includegraphics[width=\linewidth]{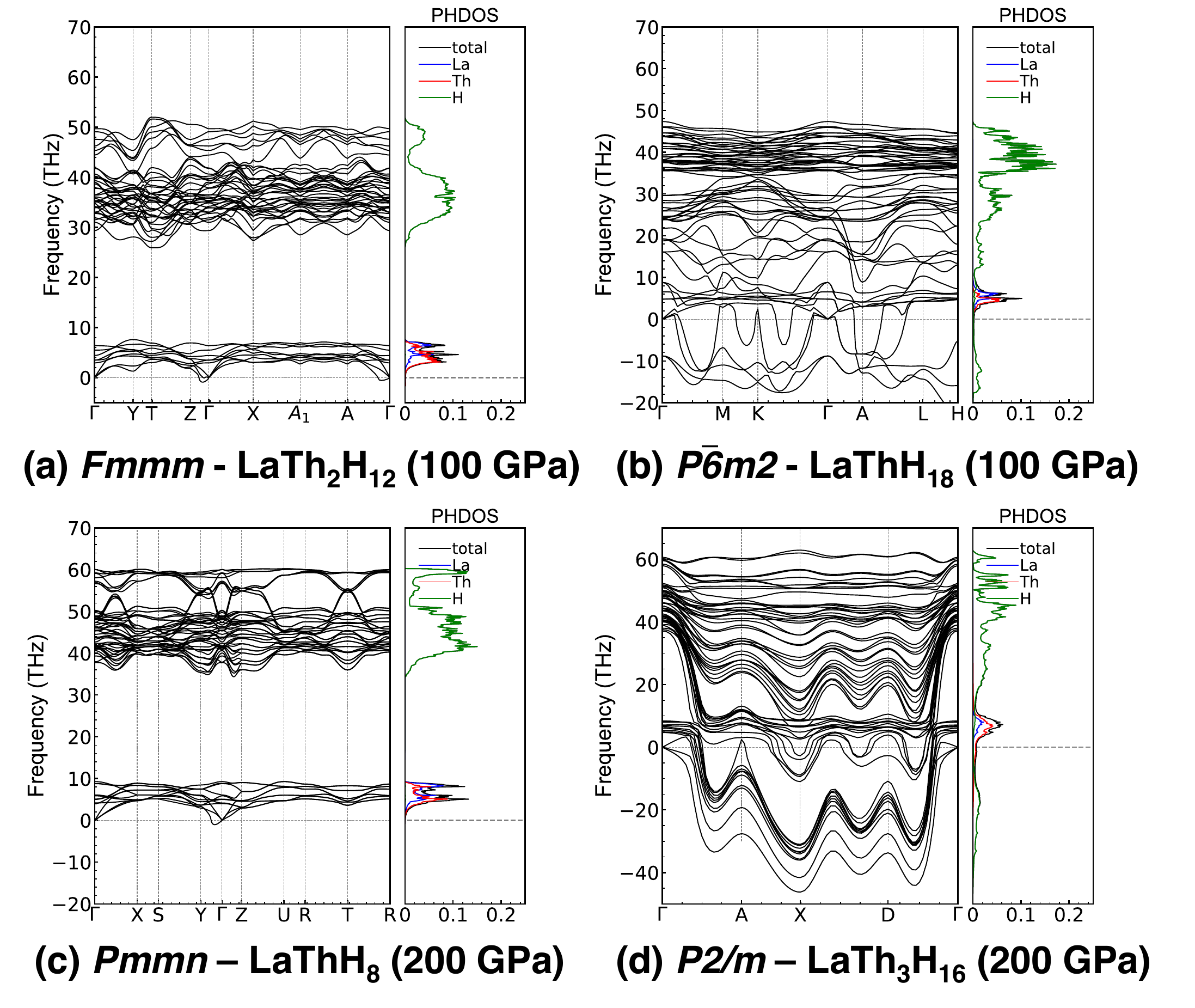}
    \caption{Phonon dispersion and phonon density of states(PHDOS) of (a) $Fmmm$-LaTh$_{2}$H$_{12}$ at 100 GPa, (b) $P\bar{6}m2$-LaThH$_{18}$ at 100 GPa, (c) $Pmmn$-LaThH$_{8}$ at 200 GPa, and (d) $P2/m$-LaTh$_{3}$H$_{16}$ at 200 GPa.}
    \label{fig.band_structure}
  \end{center}
\end{figure*}


\clearpage
\newpage
\section{Thermodynamic stability of the La-Th-H system under high pressure}

\begin{table}[htbp]
\caption{The enthalpy of formation of La$_{x}$Th$_{y}$H$_{z}$ at 5 GPa. }
\begin{tabular}{lrrrrrr}
\toprule
\hline
  Compound &  space group &  $N_\mathrm{atoms}$ &     Volume &  Enthalpy &  $\Delta H$ &  $E_\mathrm{above\_hull}$ \\
    &   &   &   \AA$^{3}$/atom &  (eV/atom) &  (eV/atom) &  (eV/atom) \\
\midrule
      H$_{2}$ &     $I4/mmm$ &       4 &   6.526 &        -3.129 &        0.000 &         0.000 \\
      La &         $Pm$ &       7 &  32.192 &        -3.727 &        0.000 &         0.000 \\
   La$_{2}$Th &       $Immm$ &       6 &  31.062 &        -4.721 &       -0.085 &         0.000 \\
    La$_{4}$H &         $P1$ &       5 &  25.586 &        -3.929 &       -0.321 &         0.000 \\
    LaH$_{2}$ & $Fm\bar{3}m$ &      12 &  14.058 &        -4.078 &       -0.750 &         0.000 \\
    LaH$_{3}$ & $Fm\bar{3}m$ &      16 &  10.179 &        -4.018 &       -0.740 &         0.000 \\
  LaThH$_{6}$ &  $Pm\bar{3}$ &       8 &   9.910 &        -4.293 &       -0.674 &         0.000 \\
      Th &   $P\bar{1}$ &       2 &  29.559 &        -6.455 &        0.000 &         0.000 \\
  Th$_{4}$H$_{15}$ & $I\bar{4}3d$ &      76 &   9.474 &        -4.397 &       -0.568 &         0.000 \\
    ThH$_{2}$ &     $I4/mmm$ &       6 &  12.833 &        -4.799 &       -0.562 &         0.000 \\
    ThH$_{3}$ &  $R\bar{3}c$ &      24 &   9.700 &        -4.550 &       -0.589 &         0.000 \\
\hline
    ThH$_{4}$ &       $C2/m$ &      10 &   9.062 &        -4.320 &       -0.526 &         0.013 \\
\hline
\bottomrule
\end{tabular}
\end{table}

\begin{table}[htbp]
\caption{The enthalpy of formation of La$_{x}$Th$_{y}$H$_{z}$ at 50 GPa. }
\begin{tabular}{lrrrrrr}
\toprule
\hline
  Compound &  space group &  $N_\mathrm{atoms}$ &     Volume &  Enthalpy &  $\Delta H$ &  $E_\mathrm{above\_hull}$ \\
    &   &   &   \AA$^{3}$/atom &  (eV/atom) &  (eV/atom) &  (eV/atom) \\
\midrule
       H$_{2}$ &      $C2/c$ &      16 &   3.004 &        -2.019 &        0.000 &         0.000 \\
       La &    $I4/mmm$ &       2 &  19.037 &         2.560 &        0.000 &         0.000 \\
     LaH$_{3}$ &    $I4/mmm$ &       8 &   7.620 &        -1.580 &       -0.706 &         0.000 \\
     LaH$_{7}$ &      $Imm2$ &      16 &   4.971 &        -1.837 &       -0.390 &         0.000 \\
   LaThH$_{6}$ & $Pm\bar{3}$ &       8 &   7.623 &        -1.887 &       -0.742 &         0.000 \\
La(ThH$_{5}$)$_{2}$ &        $Cm$ &      26 &   7.173 &        -2.036 &       -0.741 &         0.000 \\
       Th &    $I4/mmm$ &       2 &  21.154 &         0.396 &        0.000 &         0.000 \\
     ThH$_{3}$ &      $C2/c$ &      16 &   7.623 &        -2.163 &       -0.748 &         0.000 \\
     ThH$_{4}$ &      $C2/m$ &      20 &   6.490 &        -2.265 &       -0.728 &         0.000 \\
     ThH$_{6}$ &       $R3m$ &      21 &   5.369 &        -2.216 &       -0.542 &         0.000 \\
   Th$_{3}$H$_{10}$ &      $Immm$ &      26 &   7.156 &        -2.218 &       -0.756 &         0.000 \\
\hline
   LaThH$_{8}$ &      $Cmmm$ &      20 &   6.120 &        -1.963 &       -0.643 &         0.011 \\
    La$_{3}$H$_{2}$ &      $Immm$ &      10 &  13.069 &         0.355 &       -0.374 &         0.003 \\
 La$_{3}$ThH$_{16}$ &  $P\bar{1}$ &      20 &   6.111 &        -1.819 &       -0.607 &         0.009 \\
\hline
\bottomrule
\end{tabular}
\end{table}

\begin{table}[htbp]
\caption{The enthalpy of formation of La$_{x}$Th$_{y}$H$_{z}$ at 100 GPa. }
\begin{tabular}{lrrrrrr}
\toprule
\hline
  Compound &  space group &  $N_\mathrm{atoms}$ &     Volume &  Enthalpy &  $\Delta H$ &  $E_\mathrm{above\_hull}$ \\
    &   &   &   \AA$^{3}$/atom &  (eV/atom) &  (eV/atom) &  (eV/atom) \\
\midrule
     LaH$_{5}$ &   $P\bar{1}$ &      12 &   4.579 &        -0.170 &       -0.497 &         0.000 \\
   LaThH$_{8}$ &       $Cmmm$ &      20 &   5.190 &        -0.214 &       -0.689 &         0.000 \\
     LaH$_{3}$ &       $Cmcm$ &      16 &   5.962 &         0.460 &       -0.634 &         0.000 \\
       H$_{2}$ &       $C2/c$ &      24 &   2.283 &        -1.208 &        0.000 &         0.000 \\
    ThH$_{10}$ & $Fm\bar{3}m$ &      44 &   3.378 &        -0.923 &       -0.408 &         0.000 \\
     ThH$_{9}$ &         $Cm$ &      20 &   3.585 &        -0.890 &       -0.445 &         0.000 \\
     ThH$_{4}$ &       $Fmmm$ &      20 &   5.283 &        -0.445 &       -0.762 &         0.000 \\
       La &     $I4/mmm$ &       2 &  16.235 &         8.002 &        0.000 &         0.000 \\
       Th &     $I4/mmm$ &       2 &  17.738 &         6.416 &        0.000 &         0.000 \\
     LaH$_{4}$ &     $I4/mmm$ &      10 &   5.092 &         0.070 &       -0.564 &         0.000 \\
    La$_{6}$Th &       $C2/m$ &      14 &  16.409 &         7.772 &       -0.003 &         0.000 \\
    LaH$_{10}$ &         $C2$ &      22 &   3.407 &        -0.661 &       -0.291 &         0.000 \\
 LaTh$_{3}$H$_{16}$ &         $P1$ &      20 &   5.237 &        -0.336 &       -0.732 &         0.000 \\
 La$_{3}$ThH$_{16}$ &   $P\bar{1}$ &      20 &   5.148 &        -0.078 &       -0.632 &         0.000 \\
  LaThH$_{18}$ & $P\bar{6}m2$ &      20 &   3.532 &        -0.752 &       -0.385 &         0.000 \\
\hline
   La$_{3}$H$_{10}$ &       $Cmmm$ &      26 &   5.518 &         0.323 &       -0.595 &         0.012 \\
     LaH$_{9}$ &       $Imm2$ &      20 &   3.656 &        -0.595 &       -0.308 &         0.007 \\
  LaThH$_{20}$ &       $C2/m$ &      44 &   3.370 &        -0.791 &       -0.348 &         0.003 \\
La(ThH$_{6}$)$_{2}$ &       $Fmmm$ &      60 &   5.220 &        -0.295 &       -0.718 &         0.000 \\
\hline
\bottomrule
\end{tabular}
\end{table}

\begin{table}[htbp]
\caption{The enthalpy of formation of La$_{x}$Th$_{y}$H$_{z}$ at 200 GPa. }
\begin{tabular}{lrrrrrr}
\toprule
\hline
  Compound &  space group &  N$_{atoms}$ &     Volume &  Enthalpy &  $\Delta H$ &  $E_\mathrm{above\_hull}$ \\
    &   &   &   \AA$^{3}$/atom &  (eV/atom) &  (eV/atom) &  (eV/atom) \\
\midrule
    LaThH$_{8}$ &       $Pmmn$ &      20 &   4.255 &         2.701 &       -0.671 &         0.000 \\
    La$_{3}$H$_{10}$ &       $Cmmm$ &      26 &   4.501 &         3.410 &       -0.571 &         0.000 \\
        H$_{2}$ &       $C2/c$ &      24 &   1.724 &         0.017 &        0.000 &         0.000 \\
        Th &     $I4/mmm$ &       2 &  14.722 &        16.391 &        0.000 &         0.000 \\
      LaH$_{8}$ &       $C2/m$ &      36 &   3.016 &         1.548 &       -0.377 &         0.000 \\
  La$_{2}$ThH$_{12}$ &       $C2/m$ &      30 &   4.216 &         2.766 &       -0.633 &         0.000 \\
 La(ThH$_{6}$)$_{2}$ &       $C2/m$ &      30 &   4.294 &         2.638 &       -0.707 &         0.000 \\
     LaH$_{16}$ &     $P6/mmm$ &      17 &   2.360 &         0.793 &       -0.235 &         0.000 \\
      LaH$_{2}$ &       $C2/m$ &       6 &   5.700 &         5.161 &       -0.582 &         0.000 \\
      ThH$_{4}$ &     $I4/mmm$ &      10 &   4.385 &         2.541 &       -0.750 &         0.000 \\
     ThH$_{10}$ & $Fm\bar{3}m$ &      44 &   2.831 &         0.995 &       -0.510 &         0.000 \\
      ThH$_{9}$ & $P6_{3}/mmc$ &      20 &   2.987 &         1.114 &       -0.541 &         0.000 \\
  LaTh$_{3}$H$_{16}$ &       $P2/m$ &      20 &   4.317 &         2.612 &       -0.719 &         0.000 \\
      LaH$_{4}$ &     $I4/mmm$ &      10 &   4.131 &         2.909 &       -0.544 &         0.000 \\
  LaTh$_{3}$H$_{40}$ &     $I4/mmm$ &      88 &   2.814 &         1.051 &       -0.473 &         0.000 \\
        La &     $I4/mmm$ &       2 &  13.565 &        17.196 &        0.000 &         0.000 \\
  La$_{3}$ThH$_{40}$ &     $I4/mmm$ &      88 &   2.779 &         1.175 &       -0.385 &         0.000 \\
     LaH$_{10}$ & $Fm\bar{3}m$ &      44 &   2.761 &         1.241 &       -0.337 &         0.000 \\
   LaThH$_{20}$ &       $C2/m$ &      44 &   2.796 &         1.111 &       -0.431 &         0.000 \\
La(ThH$_{15}$)$_{2}$ &       $Immm$ &      66 &   2.808 &         1.071 &       -0.459 &         0.000 \\
\hline
  La$_{2}$ThH$_{10}$ &       $Cmmm$ &      26 &   4.591 &         3.275 &       -0.644 &         0.007 \\
  La$_{3}$ThH$_{16}$ &       $Pmmm$ &      20 &   4.198 &         2.804 &       -0.609 &         0.002 \\
      LaH$_{6}$ &       $C2/m$ &      28 &   3.441 &         2.040 &       -0.431 &         0.006 \\
    La$_{2}$H$_{15}$ &         $Cm$ &      34 &   3.088 &         1.655 &       -0.383 &         0.007 \\
      LaH$_{3}$ &       $Cmcm$ &      16 &   4.726 &         3.741 &       -0.570 &         0.003 \\
     LaTh$_{9}$ &       $C2/m$ &      20 &  14.592 &        16.484 &        0.013 &         0.013 \\
    La$_{7}$Th$_{3}$ &       $C2/m$ &      20 &  13.947 &        16.963 &        0.008 &         0.008 \\
   LaThH$_{18}$ & $P\bar{6}m2$ &      20 &   2.945 &         1.249 &       -0.445 &         0.010 \\
\hline
\bottomrule
\end{tabular}
\end{table}

\clearpage
\newpage
\section{Structural information of newly-predicted structures of ternary La-Th-H compounds}

\begin{center}
\begin{longtable}[htbp]{c c c c lccc}
\caption[aaaa]{Crystal structures of La$_{x}$Th$_{y}$H$_{z}$ predicted at different pressure~($P$). Lattice parameters ($a$, $b$, and $c$) are given in unit of \AA.}\\ \hline
\endfirsthead
\multicolumn{8}{c}%
{{\bfseries \tablename\ \thetable{} -- continued from previous page}} \\
\hline \multicolumn{1}{c}{Compound} &
\multicolumn{1}{c}{Space group} &
\multicolumn{1}{c}{$P$~(GPa)}&
\multicolumn{1}{c}{Lattice parameters} &
\multicolumn{4}{c}{Atomic coordinates (fractional)}\\
&&&&   \multicolumn{1}{c}{Atoms} &\multicolumn{1}{c}{$x$}& \multicolumn{1}{c}{$y$} &\multicolumn{1}{c}{$z$}\\
\hline
\endhead
\hline \multicolumn{8}{r}{{Continued on next page}} \\ \hline
\endfoot
\hline \hline
\endlastfoot
Compound & Space group & $P$~(GPa)& Lattice parameters & \multicolumn{4}{c}{Atomic coordinates (fractional)}\\
&&&&    Atoms  & $x$ &  $y$  & $z$ \\
\hline
 \hline
   \specialrule{0em}{1pt}{1pt}
LaThH$_{6}$
&  $Pm\bar{3}$
&  5
&  $a = 4.29587$
& La(1b)&0.50000&0.50000&0.50000\\
         &&
&  $b = 4.29587 $
& Th(1a)&0.00000&0.00000&0.00000\\
         &&
&  $c = 4.29587 $
& H(6g)&0.00000&0.23981&0.50000\\
         &&
& $\alpha  = 90.0 ^{\circ}$
&  & &  & \\
         &&
& $\beta  = 90.0 ^{\circ}$
&  & &  & \\
         &&
& $\gamma  = 90.0 ^{\circ}$
&  & &  & \\
\hline
\hline
   \specialrule{0em}{1pt}{1pt}
LaTh$_{2}$H$_{10}$
&  $Cm$
&  50
&  $a = 12.68599$
& La(2a)&0.14546&0.50000&0.26568\\
         &&
&  $b = 3.99464 $
& Th(2a)&0.31215&0.00000&0.93163\\
         &&
&  $c = 3.86265 $
& Th(2a)&0.48463&0.50000&0.60396\\
         &&
& $\alpha  = 90.0 ^{\circ}$
& H(4b)&0.15505&0.23494&0.77542\\
         &&
& $\beta  = 107.6919 ^{\circ}$
& H(4b)&0.46833&0.24281&0.08755\\
         &&
& $\gamma  = 90.0 ^{\circ}$
& H(2a)&0.04423&0.50000&0.66508\\
&&&& H(2a)&0.08073&0.00000&0.19966\\
&&&& H(2a)&0.23129&0.00000&0.35082\\
&&&& H(2a)&0.31640&0.50000&0.18825\\
&&&& H(2a)&0.31644&0.50000&0.68362\\
&&&& H(2a)&0.39711&0.00000&0.51624\\
\hline
\hline
   \specialrule{0em}{1pt}{1pt}
La$_{3}$ThH$_{16}$
&  $P\bar{1}$
&  100
&  $a = 3.66575$
& La(2i)&0.24898&0.00374&0.74723\\
         &&
&  $b = 5.51584 $
& La(1g)&0.00000&0.50000&0.50000\\
         &&
&  $c = 5.51636 $
& Th(1e)&0.50000&0.50000&0.00000\\
         &&
& $\alpha  = 72.3594 ^{\circ}$
& H(2i)&0.05115&0.32440&0.91432\\
         &&
& $\beta  = 75.8071 ^{\circ}$
& H(2i)&0.06142&0.62549&0.81378\\
         &&
& $\gamma  = 87.949 ^{\circ}$
& H(2i)&0.18779&0.87279&0.43907\\
&&&& H(2i)&0.19694&0.18191&0.33423\\
&&&& H(2i)&0.30500&0.81911&0.15751\\
&&&& H(2i)&0.31377&0.12601&0.06290\\
&&&& H(2i)&0.43733&0.37737&0.68941\\
&&&& H(2i)&0.44173&0.67627&0.59506\\
\hline
\hline
   \specialrule{0em}{1pt}{1pt}
LaThH$_{8}$
&  $Cmmm$
&  100
&  $a = 4.1493$
& La(2c)&0.00000&0.50000&0.50000\\
         &&
&  $b = 6.03504 $
& Th(2a)&0.00000&0.00000&0.00000\\
         &&
&  $c = 4.14484 $
& H(8m)&0.25000&0.25000&0.24643\\
         &&
& $\alpha  = 90.0 ^{\circ}$
& H(4j)&0.00000&0.14038&0.50000\\
         &&
& $\beta  = 90.0 ^{\circ}$
& H(4i)&0.00000&0.36104&0.00000\\
         &&
& $\gamma  = 90.0 ^{\circ}$
&  & &  & \\
\hline
\hline
   \specialrule{0em}{1pt}{1pt}
LaTh$_{2}$H$_{12}$
&  $Fmmm$
&  100
&  $a = 4.16354$
& La(4a)&0.00000&0.00000&0.00000\\
         &&
&  $b = 6.02118 $
& Th(8i)&0.00000&0.00000&0.33571\\
         &&
&  $c = 12.49655 $
& H(16m)&0.00000&0.13745&0.16923\\
         &&
& $\alpha  = 90.0 ^{\circ}$
& H(16j)&0.25000&0.25000&0.08443\\
         &&
& $\beta  = 90.0 ^{\circ}$
& H(8h)&0.00000&0.14072&0.50000\\
         &&
& $\gamma  = 90.0 ^{\circ}$
& H(8f)&0.25000&0.25000&0.25000\\
\hline
\hline
   \specialrule{0em}{1pt}{1pt}
La$_{1}$Th$_{3}$H$_{16}$
&  $P1$
&  100
&  $a = 4.17259$
& La(1a)&0.19998&0.12503&0.17500\\
         &&
&  $b = 5.54939 $
& Th(1a)&0.45290&0.37200&0.42199\\
         &&
&  $c = 5.55003 $
& Th(1a)&0.70001&0.62502&0.67494\\
         &&
& $\alpha  = 65.6541 ^{\circ}$
& Th(1a)&0.94702&0.87802&0.92803\\
         &&
& $\beta  = 67.9351 ^{\circ}$
& H(1a)&0.07807&0.74719&0.30201\\
         &&
& $\gamma  = 67.9394 ^{\circ}$
& H(1a)&0.07809&0.25196&0.79718\\
&&&& H(1a)&0.19996&0.76170&0.53843\\
&&&& H(1a)&0.19998&0.48818&0.81172\\
&&&& H(1a)&0.32194&0.99797&0.55299\\
&&&& H(1a)&0.32199&0.50298&0.04788\\
&&&& H(1a)&0.45012&0.01202&0.78759\\
&&&& H(1a)&0.45026&0.73750&0.06191\\
&&&& H(1a)&0.57271&0.74807&0.30104\\
&&&& H(1a)&0.57288&0.25094&0.79800\\
&&&& H(1a)&0.69991&0.26402&0.03607\\
&&&& H(1a)&0.69994&0.98619&0.31386\\
&&&& H(1a)&0.82716&0.99897&0.55202\\
&&&& H(1a)&0.82721&0.50195&0.04890\\
&&&& H(1a)&0.94977&0.23784&0.56255\\
&&&& H(1a)&0.94983&0.51246&0.28799\\
\hline
\hline
   \specialrule{0em}{1pt}{1pt}
LaThH$_{18}$
&  $P\bar{6}m2$
&  100
&  $a = 3.78023$
& La(1a)&0.00000&0.00000&0.00000\\
         &&
&  $b = 3.78023 $
& Th(1d)&0.33333&0.66667&0.50000\\
         &&
&  $c = 5.72178 $
& H(6n)&0.35855&0.17927&0.31024\\
         &&
& $\alpha  = 90.0 ^{\circ}$
& H(6n)&0.51003&0.48997&0.18933\\
         &&
& $\beta  = 90.0 ^{\circ}$
& H(2g)&0.00000&0.00000&0.40149\\
         &&
& $\gamma  = 120.0 ^{\circ}$
& H(2h)&0.33333&0.66667&0.11715\\
&&&& H(1e)&0.66667&0.33333&0.00000\\
&&&& H(1f)&0.66667&0.33333&0.50000\\
\hline
\hline
   \specialrule{0em}{1pt}{1pt}
La$_{2}$ThH$_{12}$
&  $C2/m$
&  200
&  $a = 9.60651$
& La(4i)&0.16882&0.00000&0.83631\\
         &&
&  $b = 3.77097 $
& Th(2d)&0.00000&0.50000&0.50000\\
         &&
&  $c = 3.5054 $
& H(8j)&0.16652&0.24810&0.33537\\
         &&
& $\alpha  = 90.0 ^{\circ}$
& H(4g)&0.00000&0.24551&0.00000\\
         &&
& $\beta  = 95.1755 ^{\circ}$
& H(4i)&0.05026&0.00000&0.29827\\
         &&
& $\gamma  = 90.0 ^{\circ}$
& H(4i)&0.11521&0.50000&0.03308\\
&&&& H(4i)&0.21618&0.50000&0.62998\\
\hline
\hline
   \specialrule{0em}{1pt}{1pt}
LaThH$_{8}$
&  $Pmmn$
&  200
&  $a = 2.69104$
& La(2b)&0.00000&0.50000&0.87614\\
         &&
&  $b = 5.87686 $
& Th(2b)&0.00000&0.50000&0.37309\\
         &&
&  $c = 5.37976 $
& H(4e)&0.00000&0.14722&0.37557\\
         &&
& $\alpha  = 90.0 ^{\circ}$
& H(4e)&0.00000&0.15041&0.87423\\
         &&
& $\beta  = 90.0 ^{\circ}$
& H(4e)&0.00000&0.24938&0.12515\\
         &&
& $\gamma  = 90.0 ^{\circ}$
& H(4e)&0.00000&0.25090&0.62526\\
\hline
\hline
   \specialrule{0em}{1pt}{1pt}
LaTh$_{2}$H$_{12}$
&  $C2/m$
&  200
&  $a = 9.63307$
& La(2a)&0.00000&0.00000&0.00000\\
         &&
&  $b = 3.83452 $
& Th(4i)&0.16462&0.50000&0.33039\\
         &&
&  $c = 3.49719 $
& H(8j)&0.16667&0.24705&0.83160\\
         &&
& $\alpha  = 90.0 ^{\circ}$
& H(4h)&0.00000&0.24432&0.50000\\
         &&
& $\beta  = 94.2294 ^{\circ}$
& H(4i)&0.04907&0.50000&0.80496\\
         &&
& $\gamma  = 90.0 ^{\circ}$
& H(4i)&0.11847&0.00000&0.52990\\
&&&& H(4i)&0.21575&0.00000&0.14063\\
\hline
\hline
   \specialrule{0em}{1pt}{1pt}
LaTh$_{3}$H$_{16}$
&  $P2/m$
&  200
&  $a = 3.49773$
& La(1d)&0.50000&0.00000&0.00000\\
         &&
&  $b = 3.84416 $
& Th(2n)&0.25160&0.50000&0.24517\\
         &&
&  $c = 6.463 $
& Th(1c)&0.00000&0.00000&0.50000\\
         &&
& $\alpha  = 90.0 ^{\circ}$
& H(4o)&0.24791&0.24994&0.74980\\
         &&
& $\beta  = 96.44 ^{\circ}$
& H(2i)&0.00000&0.24276&0.00000\\
         &&
& $\gamma  = 90.0 ^{\circ}$
& H(2m)&0.03167&0.00000&0.17803\\
&&&& H(2n)&0.21600&0.50000&0.57161\\
&&&& H(2n)&0.28166&0.50000&0.92659\\
&&&& H(2m)&0.46750&0.00000&0.32344\\
&&&& H(2l)&0.50000&0.25574&0.50000\\
\hline
\hline
   \specialrule{0em}{1pt}{1pt}
La$_{3}$ThH$_{40}$
&  $I4/mmm$
&  200
&  $a = 4.96324$
& La(4d)&0.00000&0.50000&0.25000\\
         &&
&  $b = 4.96324 $
& La(2a)&0.00000&0.00000&0.00000\\
         &&
&  $c = 9.92467 $
& Th(2b)&0.00000&0.00000&0.50000\\
         &&
& $\alpha  = 90.0 ^{\circ}$
& H(32o)&0.12168&0.37777&0.06045\\
         &&
& $\beta  = 90.0 ^{\circ}$
& H(16m)&0.12115&0.12115&0.31088\\
         &&
& $\gamma  = 90.0 ^{\circ}$
& H(16m)&0.12145&0.12145&0.18982\\
&&&& H(16m)&0.24633&0.24633&0.87369\\
\hline
\hline
   \specialrule{0em}{1pt}{1pt}
LaThH$_{20}$
&  $R\bar{3}m$
&  200
&  $a = 3.51862$
& La(3b)&-0.00000&-0.00000&0.50000\\
         &&
&  $b = 3.51862 $
& Th(3a)&0.00000&0.00000&0.00000\\
         &&
&  $c = 17.20948 $
& H(18h)&0.00604&0.50302&0.43706\\
         &&
& $\alpha  = 90.0 ^{\circ}$
& H(18h)&0.00961&0.50480&0.93701\\
         &&
& $\beta  = 90.0 ^{\circ}$
& H(6c)&0.00000&0.00000&0.12679\\
         &&
& $\gamma  = 120.0 ^{\circ}$
& H(6c)&0.00000&0.00000&0.18941\\
&&&& H(6c)&0.00000&0.00000&0.31095\\
&&&& H(6c)&0.00000&0.00000&0.37656\\
\hline
\hline
   \specialrule{0em}{1pt}{1pt}
LaTh$_{2}$H$_{30}$
&  $Immm$
&  200
&  $a = 3.52098$
& La(2d)&0.00000&0.50000&0.00000\\
         &&
&  $b = 4.97927 $
& Th(4j)&0.00000&0.50000&0.33375\\
         &&
&  $c = 10.57092 $
& H(16o)&0.24417&0.12153&0.33314\\
         &&
& $\alpha  = 90.0 ^{\circ}$
& H(8l)&0.00000&0.12139&0.41499\\
         &&
& $\beta  = 90.0 ^{\circ}$
& H(8l)&0.00000&0.12213&0.25109\\
         &&
& $\gamma  = 90.0 ^{\circ}$
& H(8l)&0.00000&0.12435&0.91919\\
&&&& H(8l)&0.00000&0.25334&0.16495\\
&&&& H(8n)&0.24469&0.12292&0.00000\\
&&&& H(4h)&0.00000&0.24449&0.50000\\
\hline
\hline
   \specialrule{0em}{1pt}{1pt}
LaTh$_{3}$H$_{40}$
&  $I4/mmm$
&  200
&  $a = 4.98318$
& La(2b)&0.00000&0.00000&0.50000\\
         &&
&  $b = 4.98318 $
& Th(4d)&0.00000&0.50000&0.25000\\
         &&
&  $c = 9.972 $
& Th(2a)&0.00000&0.00000&0.00000\\
         &&
& $\alpha  = 90.0 ^{\circ}$
& H(32o)&0.12161&0.37692&0.43864\\
         &&
& $\beta  = 90.0 ^{\circ}$
& H(16m)&0.12240&0.12240&0.81198\\
         &&
& $\gamma  = 90.0 ^{\circ}$
& H(16m)&0.12248&0.12248&0.68885\\
&&&& H(16m)&0.24664&0.24664&0.62309\\
\hline

\end{longtable}
\end{center}

\clearpage